\documentclass[a4paper,11pt]{article}
\usepackage{pos}

\newcommand{\HEPfit}{\texttt{HEPfit}}

\title{Bayesian analyses of the A2HDM with low-mass scalars}

\author*[a]{Antonio M. Coutinho}
\emailAdd{antonio.coutinho@ific.uv.es} 

\author[a]{Anirban Karan}
\emailAdd{kanirban@ific.uv.es} 

\author[b]{V\'ictor Miralles,}
\emailAdd{victor.miralles@manchester.ac.uk}

\author[a]{Antonio Pich}
\emailAdd{antonio.pich@ific.uv.es}

\affiliation[a]{Instituto de F\'isica Corpuscular, Universitat de València -- CSIC, Parque Cient\'ifico, Catedr\'atico Jos\'e Beltr\'an 2, E-46980 Paterna, Spain}

\affiliation[b]{Department of Physics and Astronomy, University of Manchester, Oxford Road, Manchester M13 9PL, United
Kingdom}

\abstract{Two-Higgs-doublet models come with an augmented parameter space which allows them to possibly solve some of the shortcomings of the Standard Model, and opens the window to a plethora of new phenomena to be discovered. The introduction of scalar-mediated tree-level flavour-changing neutral currents may be tackled with the imposition of extra symmetries on the model or, alternatively, by demanding a strict proportionality between the flavour-changing couplings and fermion mass matrices. The latter is the very idea behind the Aligned-Two-Higgs-Doublet Model (A2HDM). The coefficients that govern such proportionality are, in general, complex and, therefore, possible new sources of CP violation, a calling card of this class of models. We present here the results of new state-of-the-art analyses of the A2HDM where, in particular, we ascertain whether current data allows the A2HDM to accommodate extra scalars lighter than the 125 GeV Higgs boson. To this effect, we make use of theoretical constraints, bounds from Higgs searches at the LHC and LEP, electroweak precision observables, and a set of flavour observables, all globally combined within \HEPfit{}, a software with a Bayesian Markov Chain Monte Carlo approach to statistical inference. Focusing on the light-pseudoscalar scenario, we find a region of parameter space compatible with all the constraints we impose.}

\FullConference{12th Large Hadron Collider Physics Conference  (LHCP2024)\\
3-7 June 2024 \\
Boston, USA \\}

\begin{document}
\maketitle

\section{The Aligned-Two-Higgs-Doublet Model}
One of the simplest extensions to the Standard Model is the Two-Higgs-Doublet Model~\cite{Branco:2011iw}, where a second Higgs doublet is added, and every term that contains the current Higgs gets duplicated. Without loss of generality, one can always rotate the two doublets into a basis where only one acquires a vacuum expectation value (VEV). In this so-called \textit{Higgs basis}, the linear parametrisation of the fields around the vacuum may be written as
\begin{equation}
H_1=\begin{pmatrix}
G^+\\
\big[v + S_1 + i\, G^0\big] / \sqrt{2}
\end{pmatrix}\, ,\quad
H_2=\begin{pmatrix}
H^+\\
\big[S_2 + i\, S_3\big] / \sqrt{2}
\end{pmatrix} \, ,
\end{equation}
and the most general renormalisable scalar potential takes the form:
\begin{align}
V&=\mu_1\,H_1^\dagger H_1+\mu_2\,H_2^\dagger H_2 +\big[\mu_3\,H_1^\dagger H_2+h.c.\big]
+\frac{\lambda_1}{2}\,(H_1^\dagger H_1)^2+\frac{\lambda_2}{2}\,(H_2^\dagger H_2)^2
+\lambda_3\,(H_1^\dagger H_1)(H_2^\dagger H_2) \nonumber\\
&+\lambda_4\,(H_1^\dagger H_2)(H_2^\dagger H_1)
+\Big[\Big(\frac{\lambda_5}{2}\,H_1^\dagger H_2+\lambda_6 \,H_1^\dagger H_1 +\lambda_7 \,H_2^\dagger H_2\Big)(H_1^\dagger H_2)+ \mathrm{h.c.}\Big] \, .
\end{align}
In the Yukawa sector, since one associates mass matrices with the the couplings to the doublet containing the VEV -- and the Goldstone bosons, $\lbrace G^0, G^{\pm} \rbrace$, which give mass to the $Z$ and $W^{\pm}$ bosons --, the extra terms stem from $H_2$,
\begin{equation}
\mathcal{L}_{Y} = -\frac{\sqrt 2}{v} \Big[
\overline{Q}_L \big( M_u\, \tilde{H}_1+N_u\, \tilde{H}_2 \big) u_R
+\overline{Q}_L \big( M_d\, H_1 + N_d\, H_2 \big) d_R
+\overline{L}_L \big( M_\ell\, H_1 + N_\ell\, H_2 \big) \ell_R
\, + \, \textrm{h.c.}\Big]
\end{equation}
and carry with them a completely arbitrary, $3\times 3$ complex matrix for each flavour, $N_f$. As such, the model introduces tree-level flavour-changing neutral currents mediated by $S_2$ and $S_3$, with these contributions being expected to enhance some flavour-changing phenomena beyond current experimental bounds. Among many proposed solutions, there is the possibility of $N_f = \varsigma_f \, M_f$, such that the diagonalisation of the mass matrices guarantees the zeroing of the new off-diagonal terms, the very idea behind the Aligned-Two-Higgs-Doublet Model (A2HDM)~\cite{Pich:2009sp}. The coefficients that govern the alignment, $\varsigma_f$, can be complex, and thus possible sources of CP violation, and are, in general, sources of added complexity in comparison to the solutions which just get rid of these interactions altogether, usually with the impostion of a $Z_2$ symmetry on the model \cite{Glashow:1976nt,Paschos:1976ay}.

Working with the CP-conserving version of the A2HDM, where all parameters of the potential and the $\varsigma_f$ are taken real, and where $S_1$ and $S_2$ can be rotated into two CP-even physical eigenstates, and $S_3$ is already a physical CP-odd state,
\begin{equation}
\begin{pmatrix}
h\\
H
\end{pmatrix}=\begin{pmatrix}
\cos\tilde{\alpha}& \sin\tilde{\alpha}\\ -\sin\tilde{\alpha}&\cos\tilde{\alpha}
\end{pmatrix}\begin{pmatrix}
S_1\\
S_2
\end{pmatrix}, \quad \text{and} \quad A = S_3\, ,
\end{equation}
the scalar-fermion couplings contained in the Yukawa Lagrangian can be cast as:
\begin{equation}
-\mathcal L_Y \subset \sum_{i,f}\frac{y_f^{\varphi^0_i}}{v}\,
\varphi^0_i \Big[\bar f M_f \mathcal{P}_R f\Big]
\, + \, \frac{\sqrt 2}{v}\, H^+ \Big[
\bar u\,\big\{\varsigma_d V M_d \mathcal{P}_R
-\varsigma_u M_u^\dagger V\mathcal{P}_L\big\}\, d
+\varsigma_\ell \bar \nu M_\ell \mathcal P_R \ell\Big]
\, + \, \mathrm{h.c.} \, ,
\end{equation}
where, in terms of the three physical neutral scalars,
\begin{equation}
y_{f}^h=\cos\tilde\alpha+\varsigma_{f}\,\sin \tilde\alpha\, , 
\quad
y_{f}^H=-\sin\tilde\alpha+\varsigma_{f}\,\cos \tilde\alpha\, ,
\quad
y_{u}^A= -i\, \varsigma_{u}\, ,
\quad
y_{d,\ell}^A= i\, \varsigma_{d,\ell}\, .
\end{equation}

Finally, a word must be said concerning the fact that, though radiative corrections introduce misalignment at higher loops, this effect is small and below the current experimental reach~\cite{Braeuninger:2010td,Jung:2010ik,Penuelas:2017ikk}.

\clearpage

\begin{table}[t!]
        \centering
        \begin{tabular}{ccccc}
        \hline\hline
        $M_A$~(GeV) & $M_H$~(GeV) & $M_{H^\pm}$~(GeV) & $\lambda_2$ & $\lambda_3$ \\
        \hline
        $(20,120)$ & $(130,700)$ & $(160,700)$ & $(-0.5,9.0)$ & $(-1,15)$ \\
        \hline\hline
        $\tilde\alpha$ & $\varsigma_u$ & $\varsigma_d$ & $\varsigma_\ell$ & $\lambda_7$ \\
        \hline
        $(-0.08,0.08)$ & $(-1,1)$ & $(-25,25)$ & $(-100,100)$ & $(-3.5,3.5)$\\
        \hline\hline
        \end{tabular}
        \caption{Ranges of the flat prior distributions chosen for the ten independent model parameters.}
        \label{tab:priors}
\end{table}

\begin{figure}[ht]
  \centering
  \scalebox{1.15}[1.0]{
  \hspace{-5mm}
  \begin{tabular}{c|c|c|c}
    \hline\hline
    Direct searches & Signal strengths & Flavour obs. & Global fit
    \\
    \hline\hline
    & & & \\
    \includegraphics[width=0.20\textwidth]{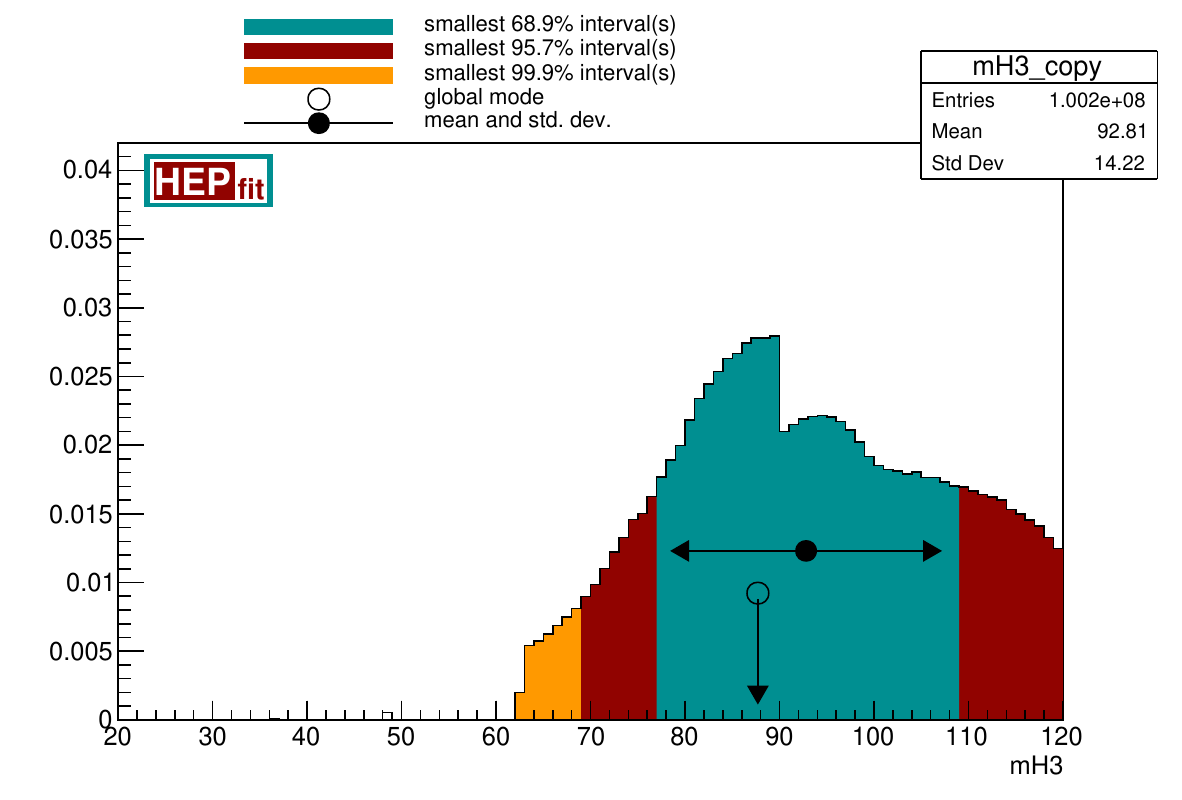}&
    \includegraphics[width=0.20\textwidth]{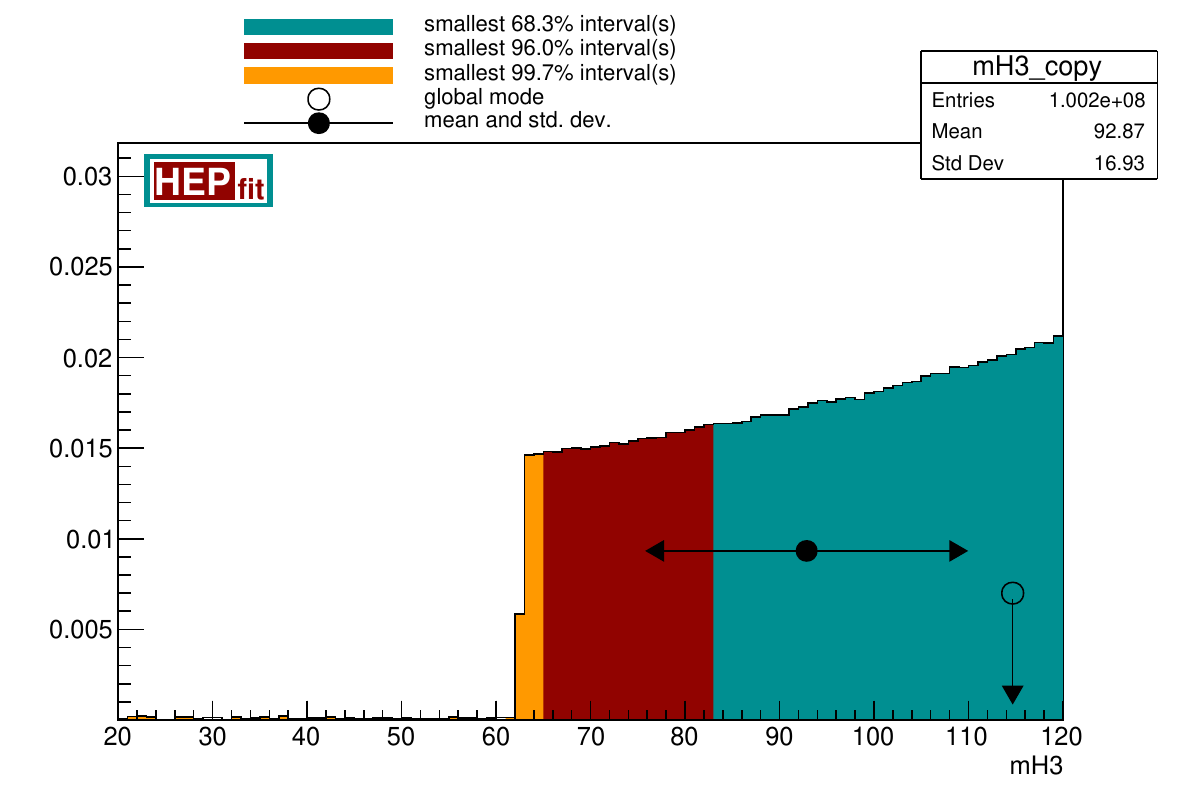}&
    \includegraphics[width=0.20\textwidth]{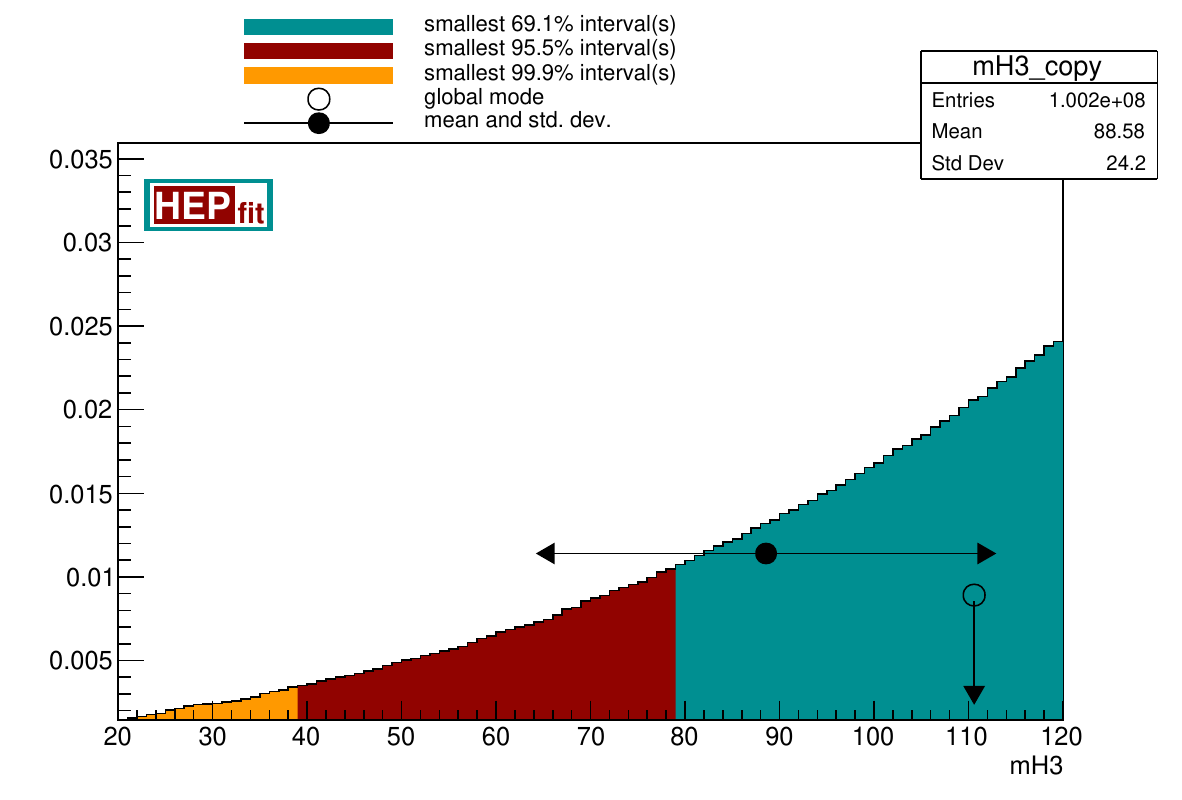}&
    \includegraphics[width=0.20\textwidth]{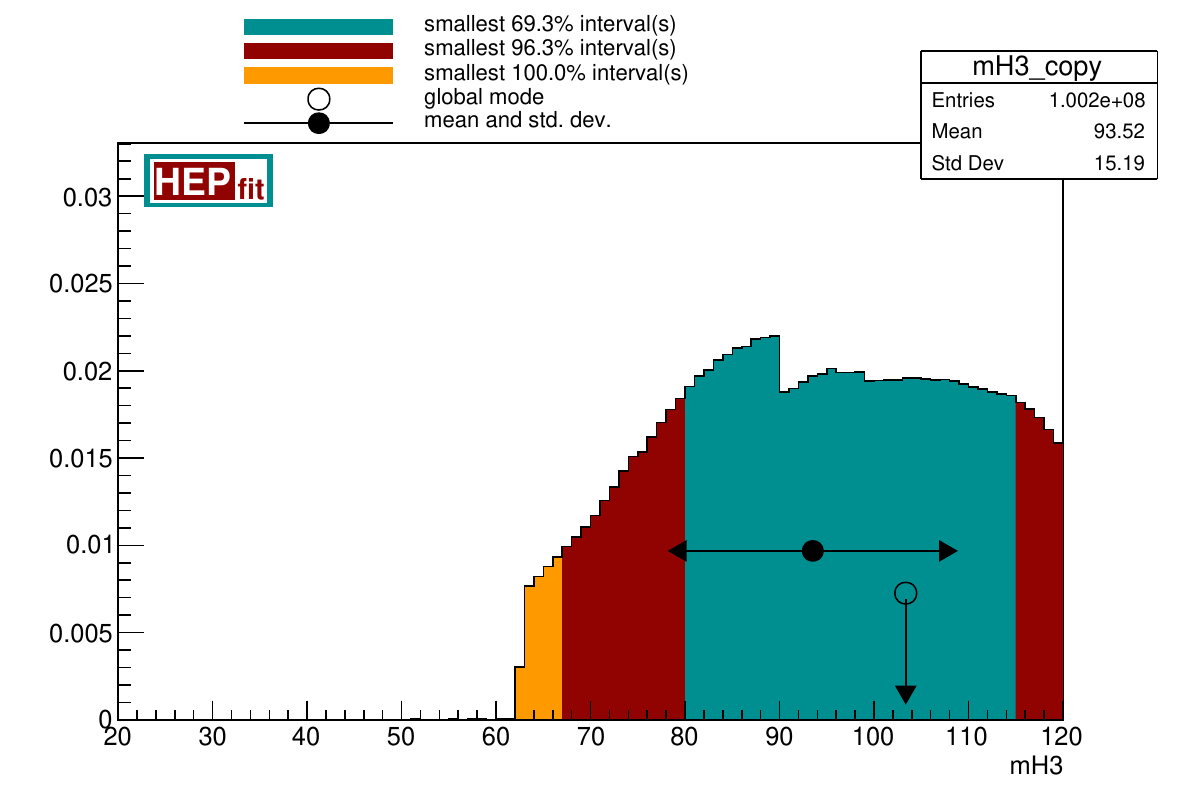}
    \\
    \includegraphics[width=0.20\textwidth]{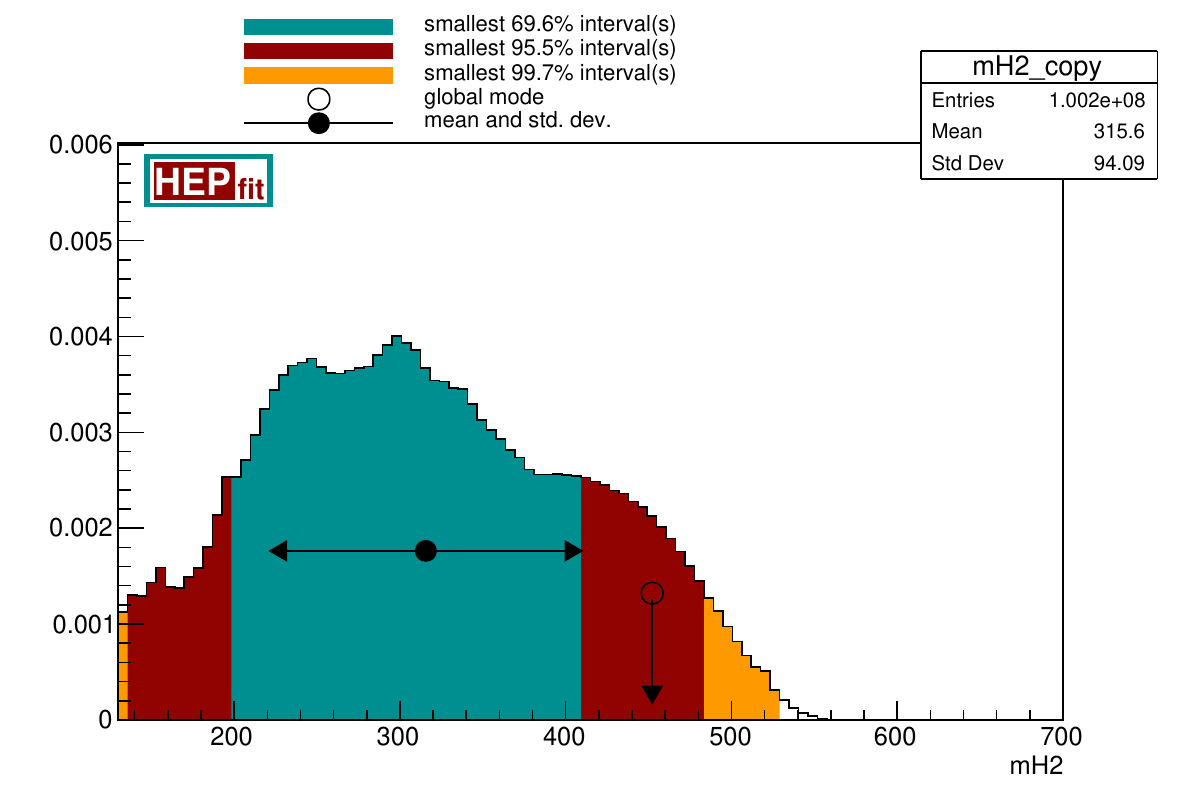}&
    \includegraphics[width=0.20\textwidth]{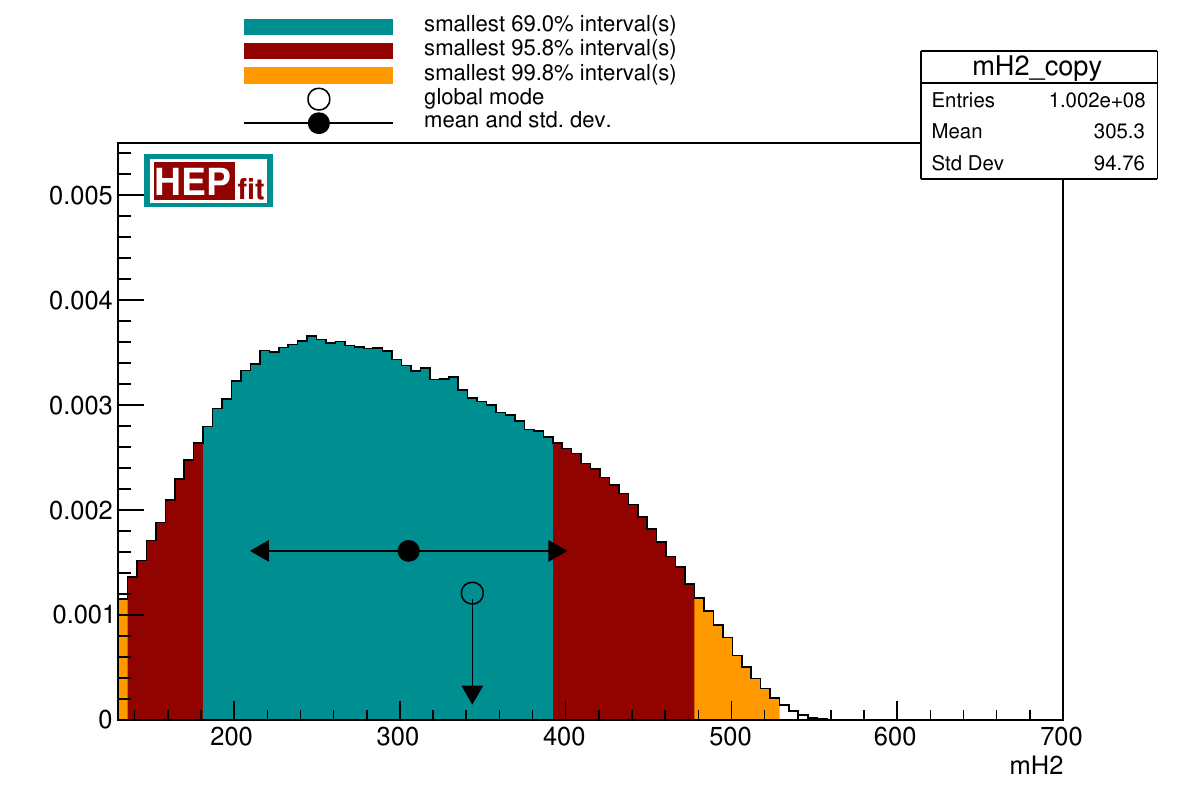}&
    \includegraphics[width=0.20\textwidth]{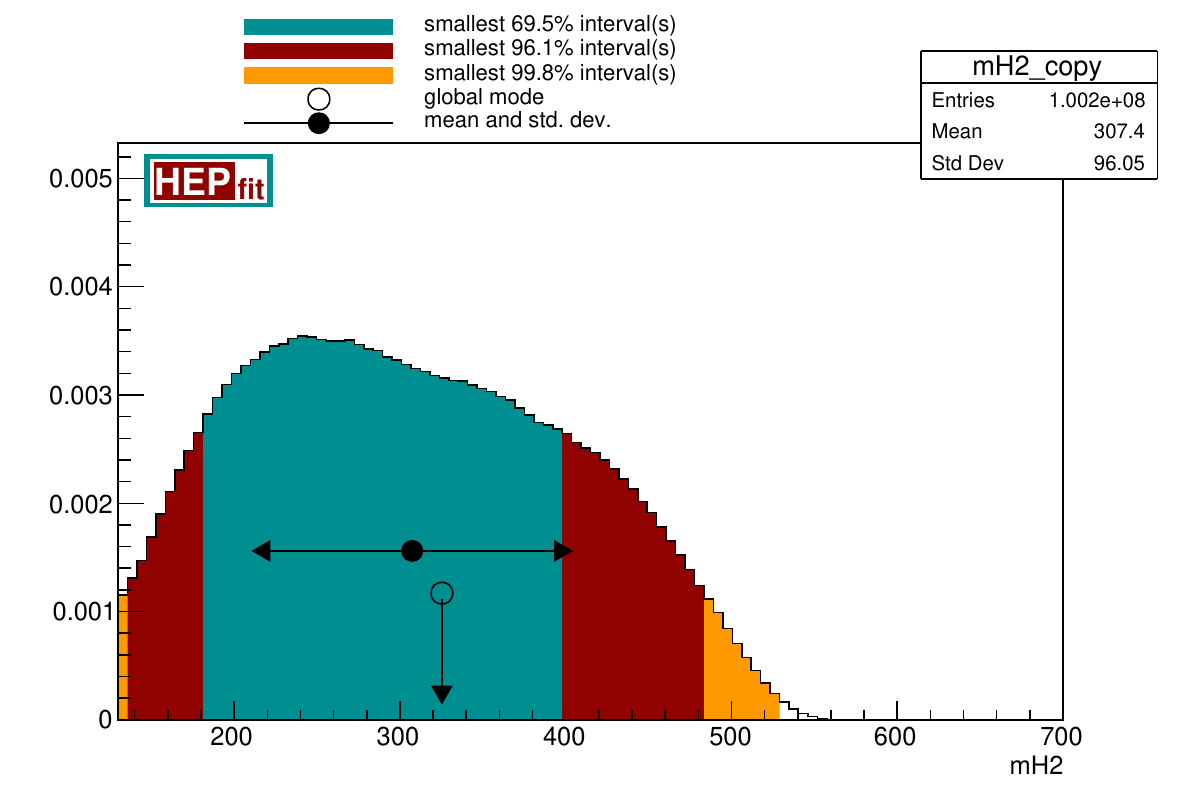}&
    \includegraphics[width=0.20\textwidth]{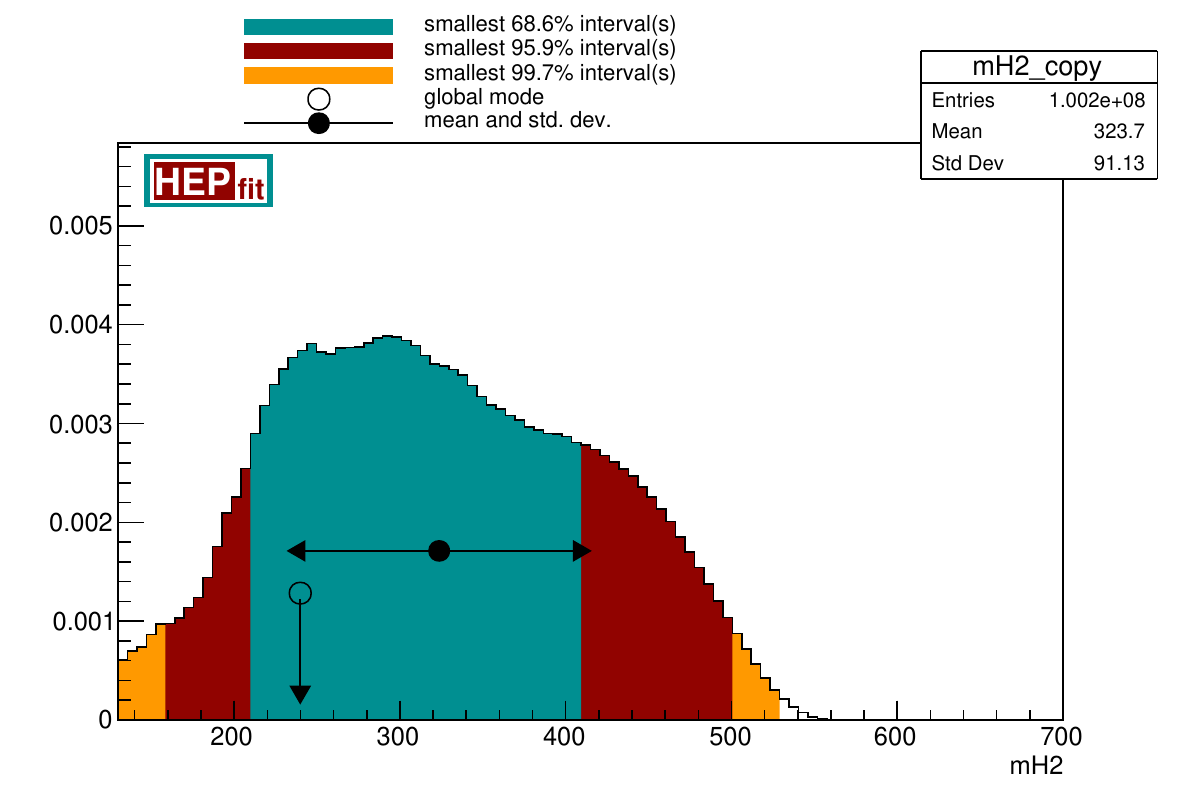}
    \\
    \includegraphics[width=0.20\textwidth]{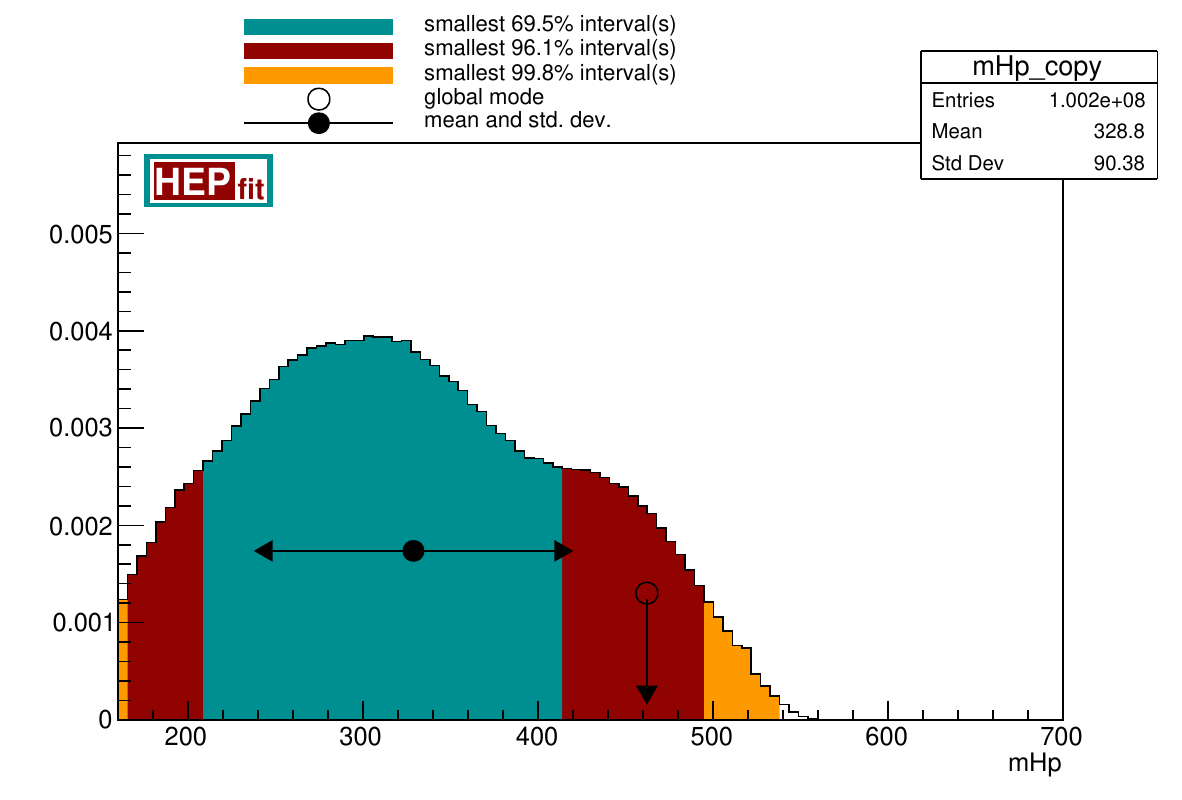}&
    \includegraphics[width=0.20\textwidth]{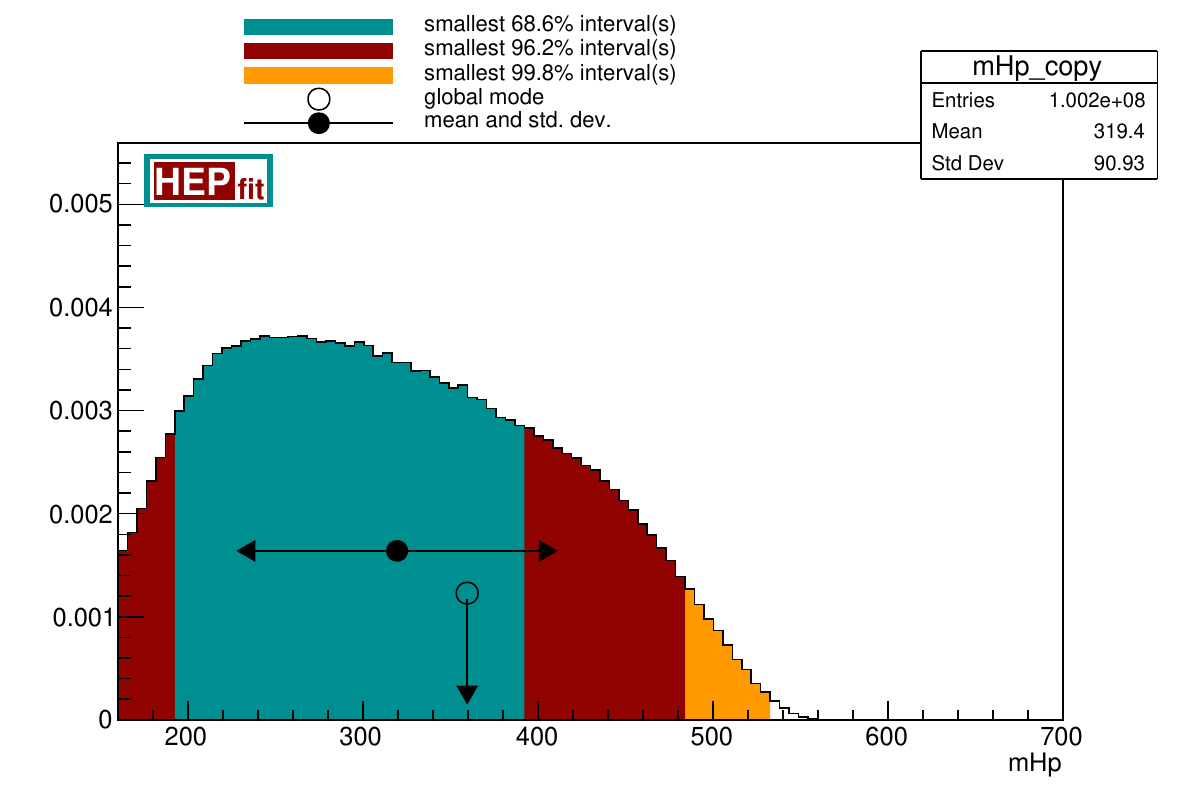}&
    \includegraphics[width=0.20\textwidth]{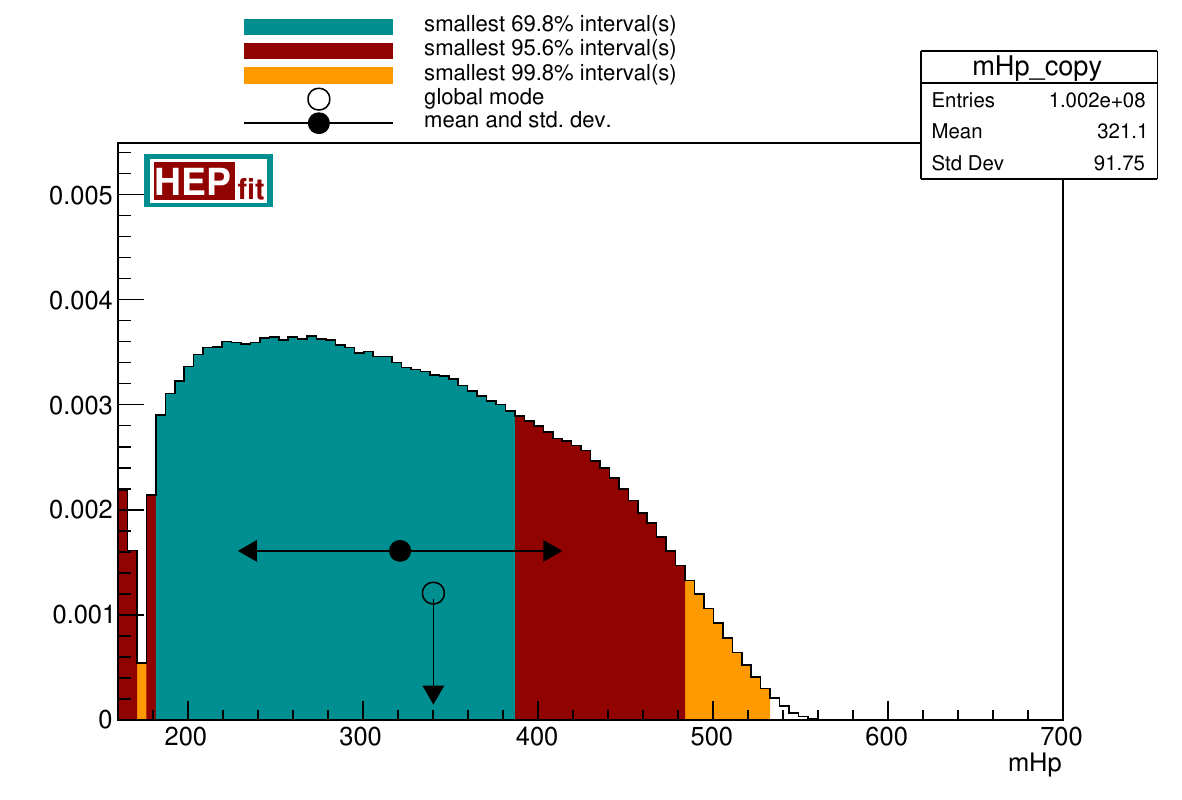}&
    \includegraphics[width=0.20\textwidth]{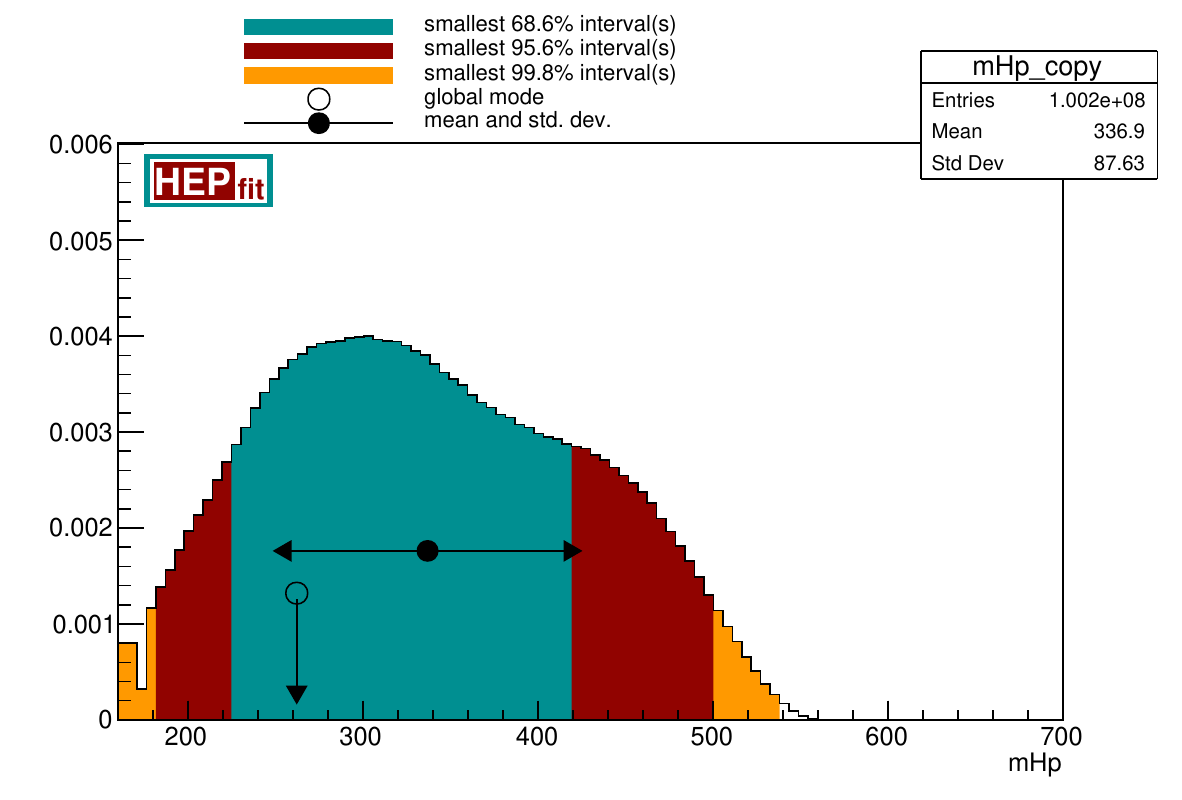}
    \\
    \includegraphics[width=0.20\textwidth]{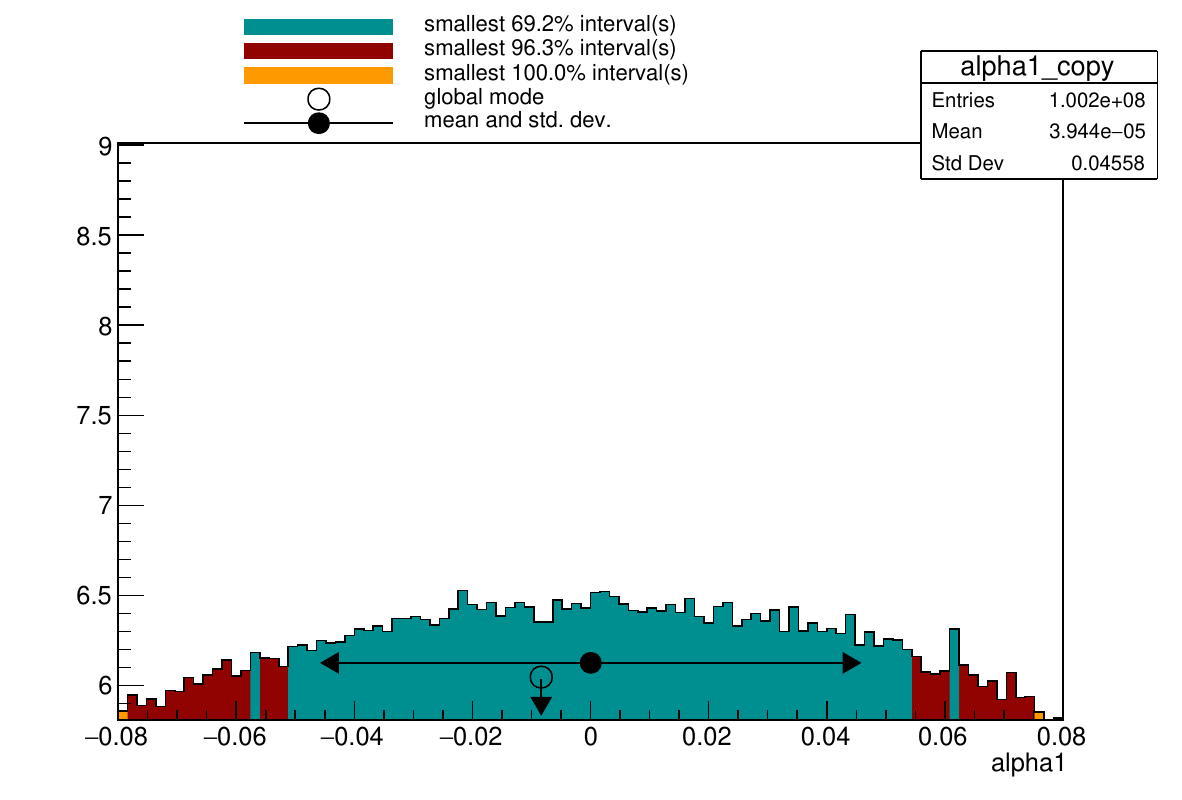}&
    \includegraphics[width=0.20\textwidth]{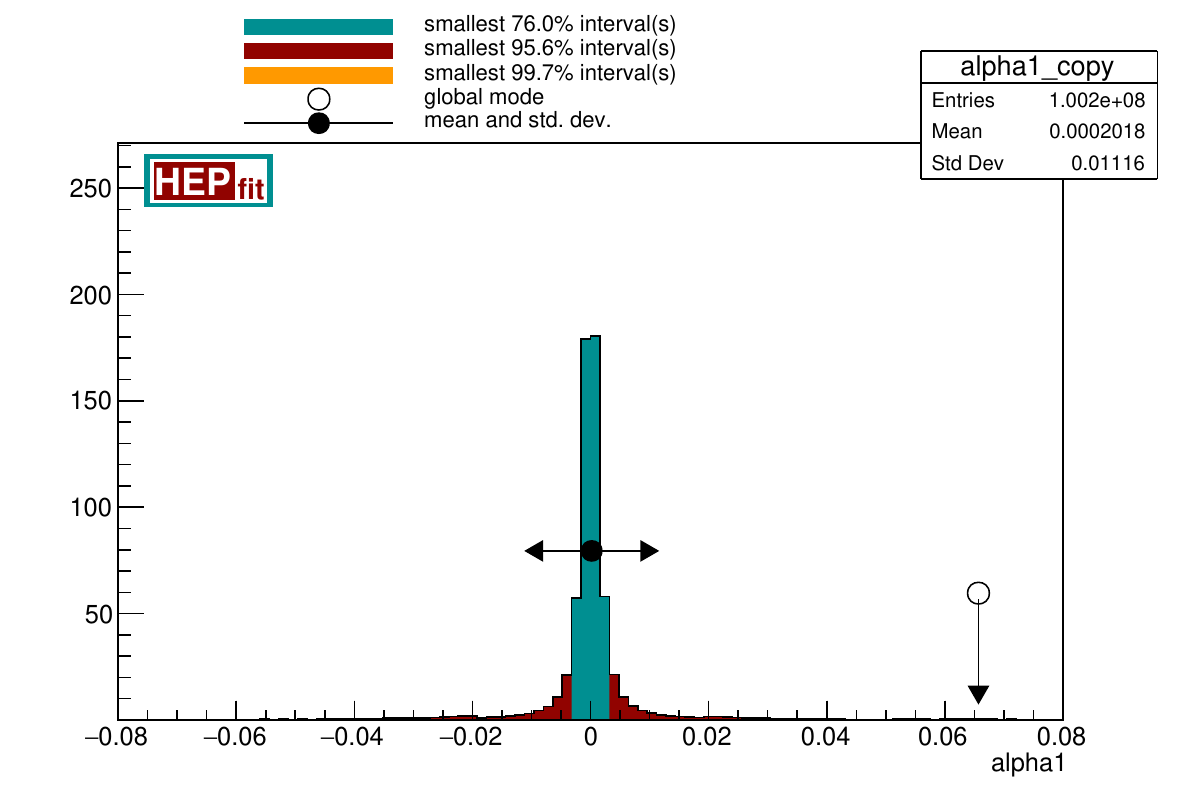}&
    \includegraphics[width=0.20\textwidth]{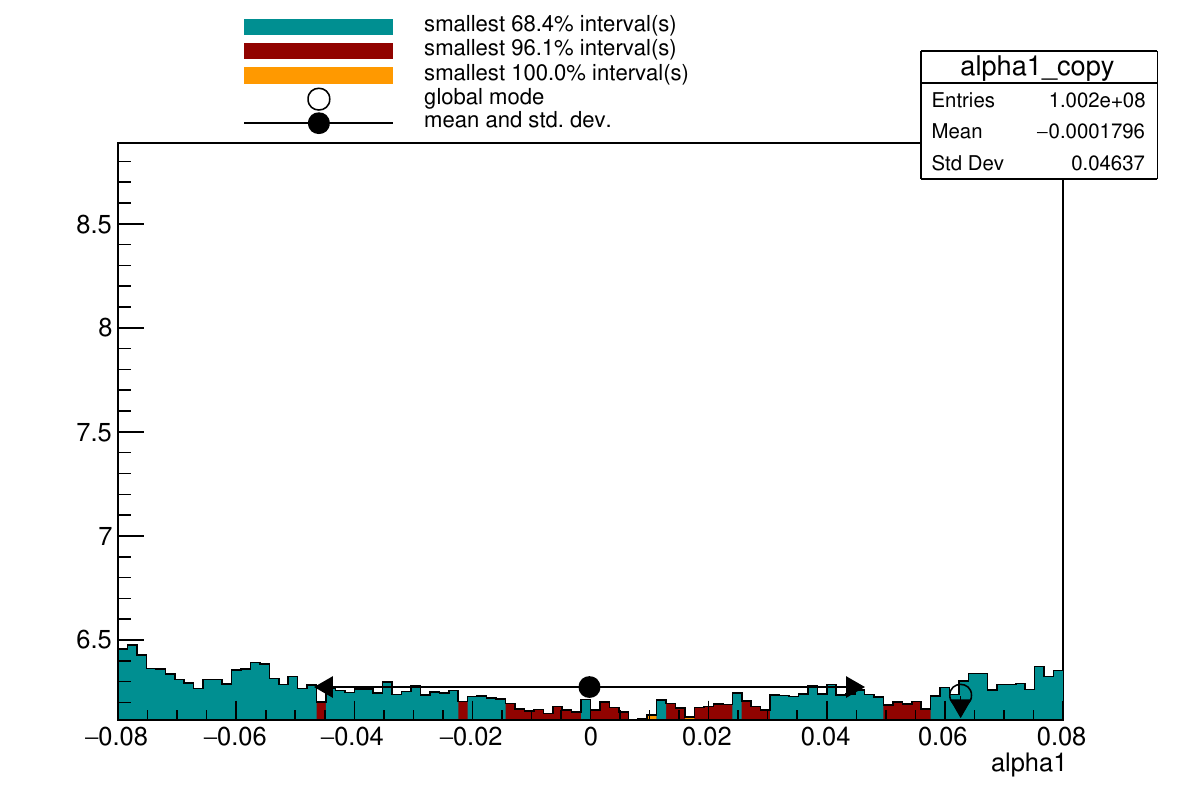}&
    \includegraphics[width=0.20\textwidth]{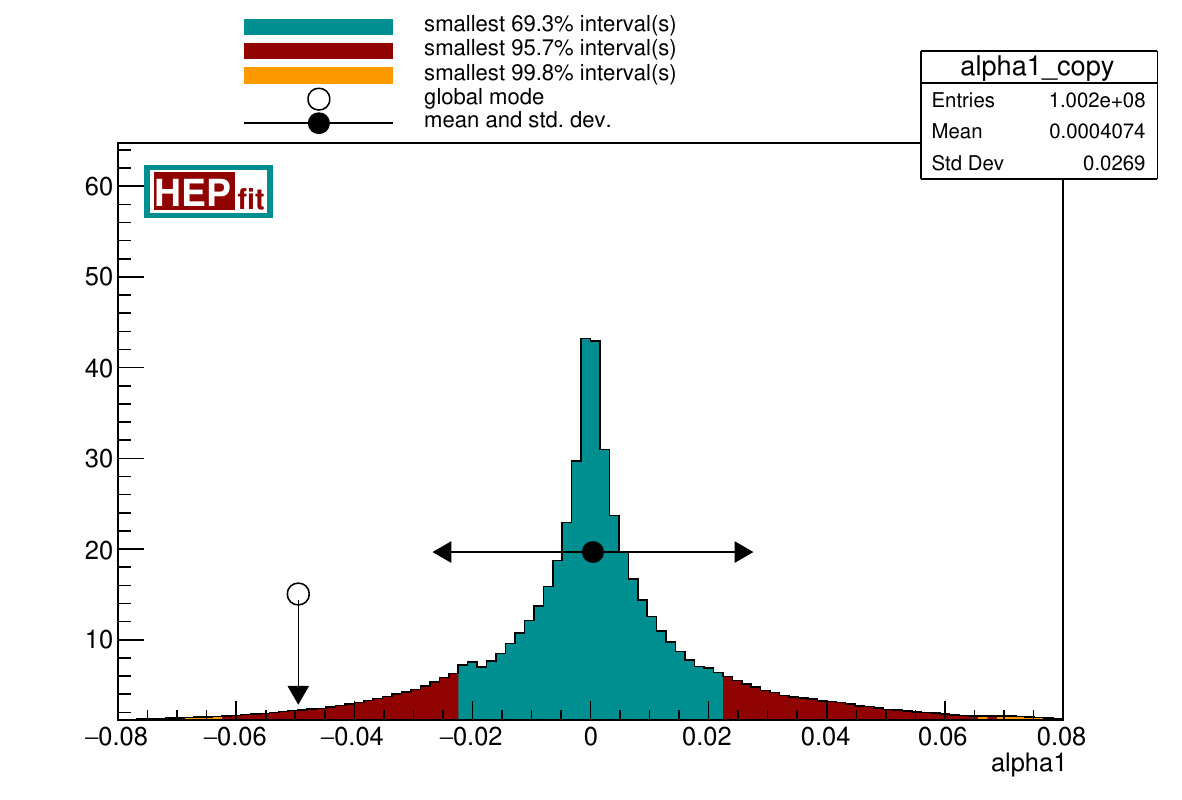}
    \\
    \includegraphics[width=0.20\textwidth]{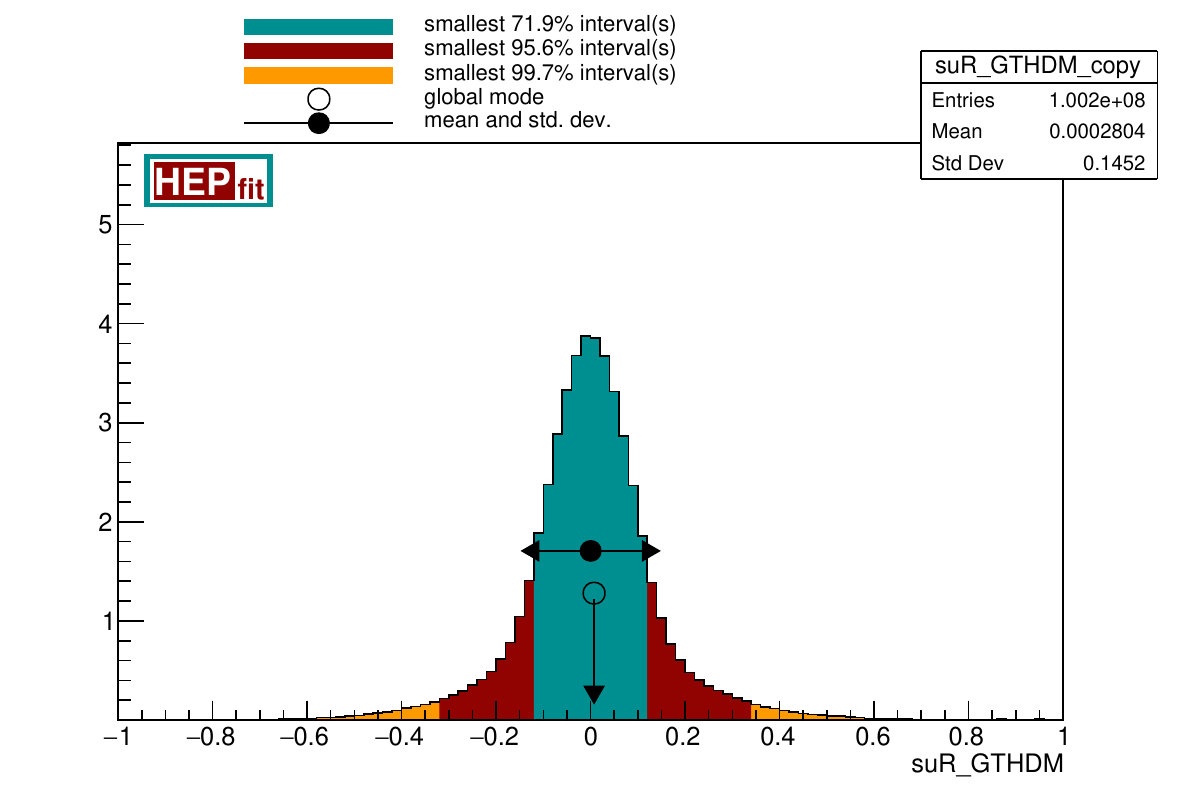}&
    \includegraphics[width=0.20\textwidth]{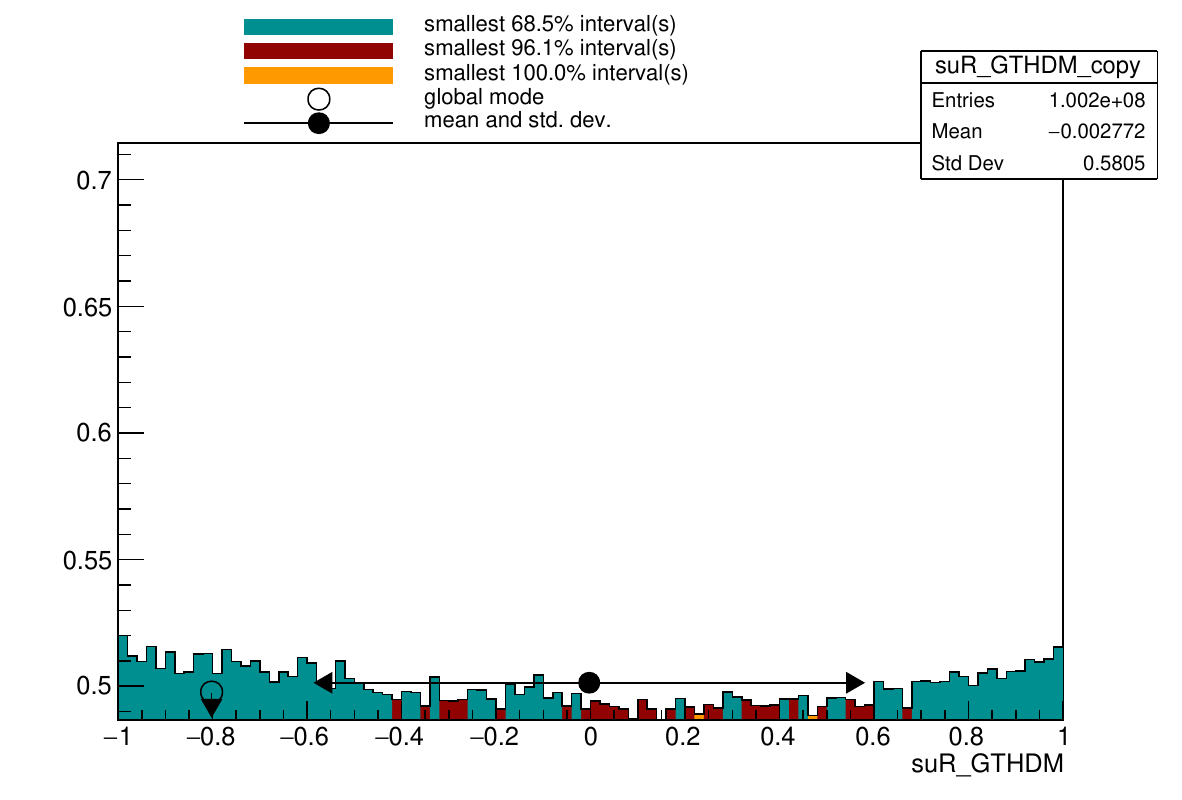}&
    \includegraphics[width=0.20\textwidth]{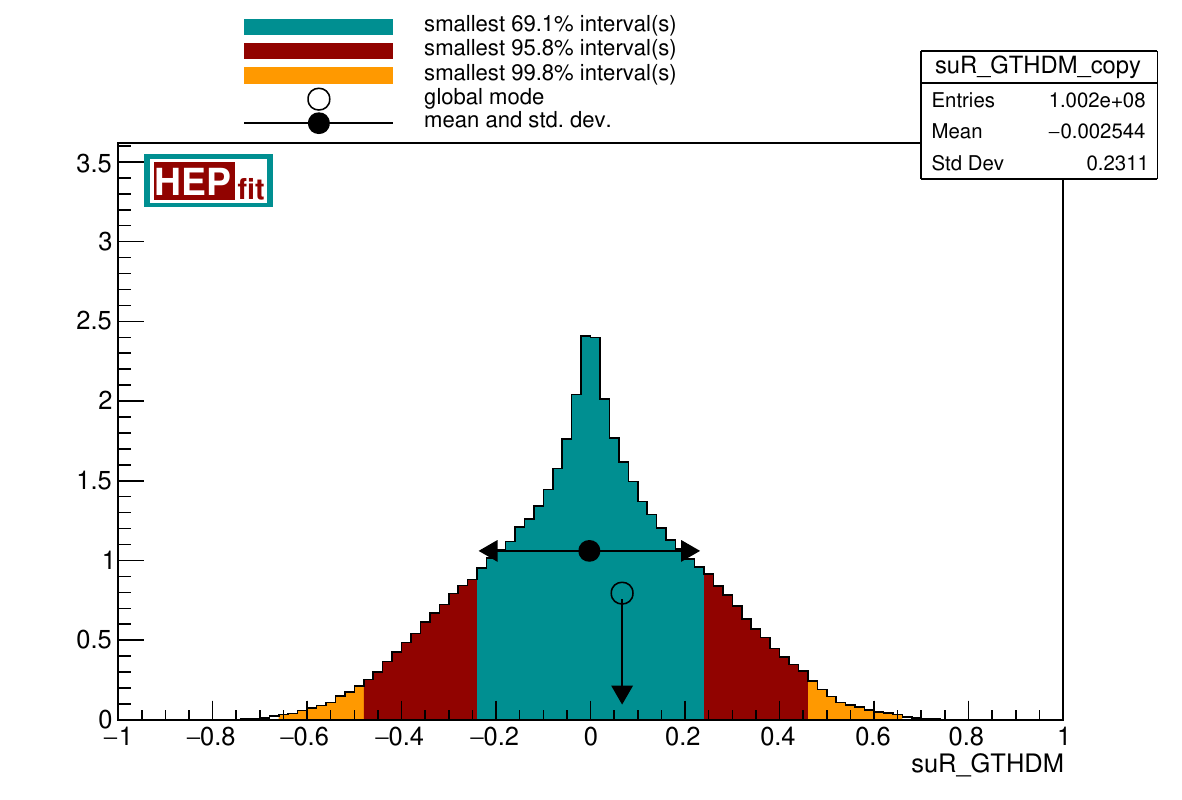}&
    \includegraphics[width=0.20\textwidth]{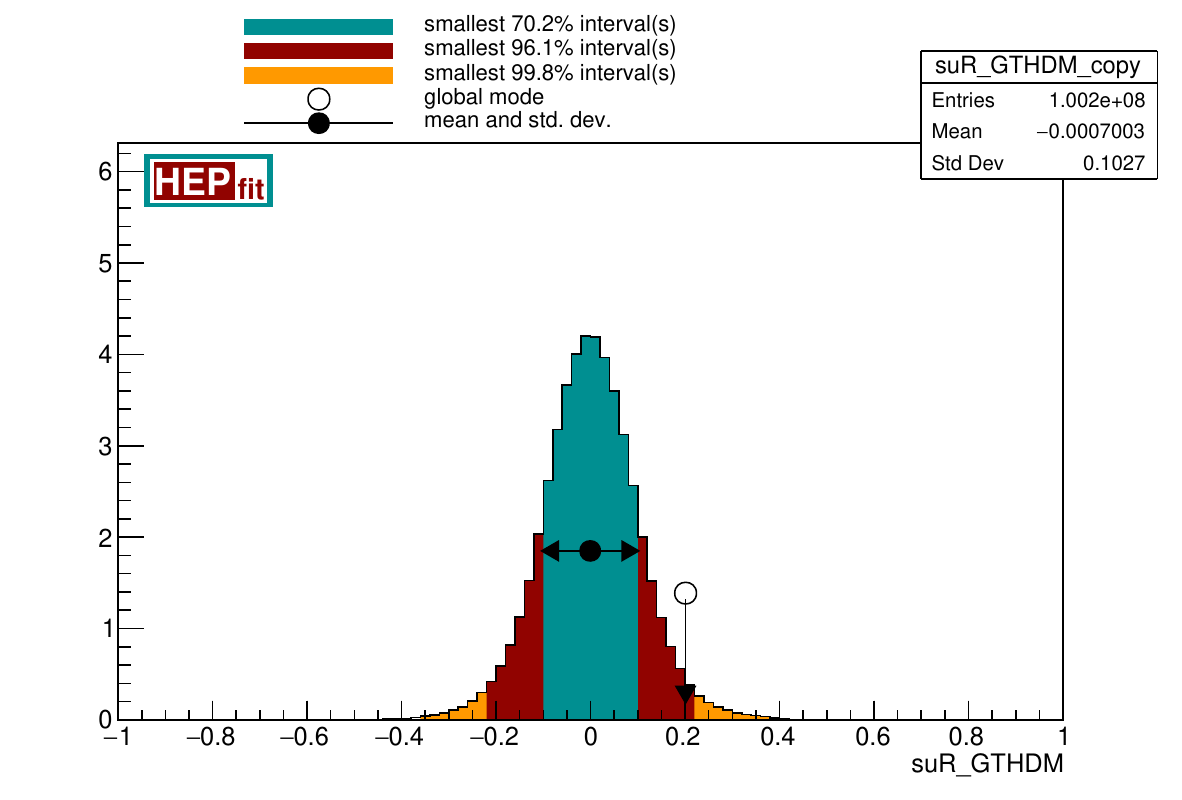}
    \\
    \includegraphics[width=0.20\textwidth]{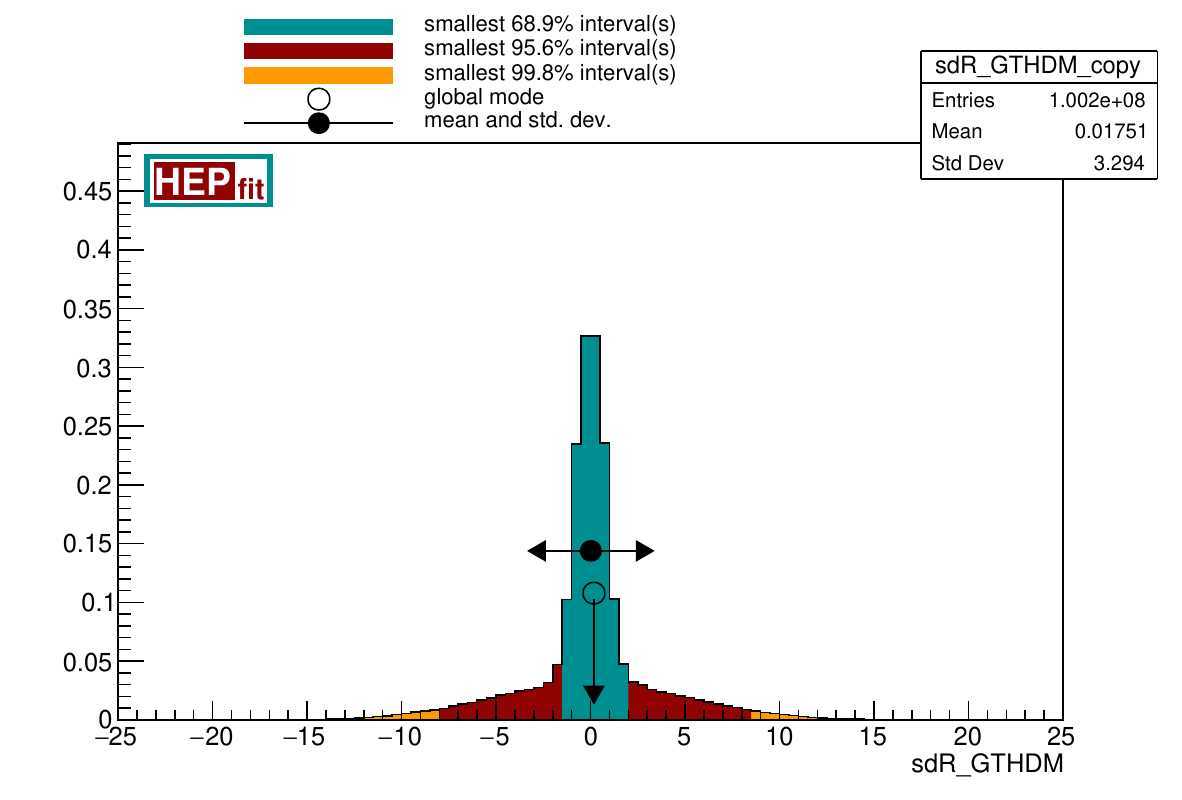}&
    \includegraphics[width=0.20\textwidth]{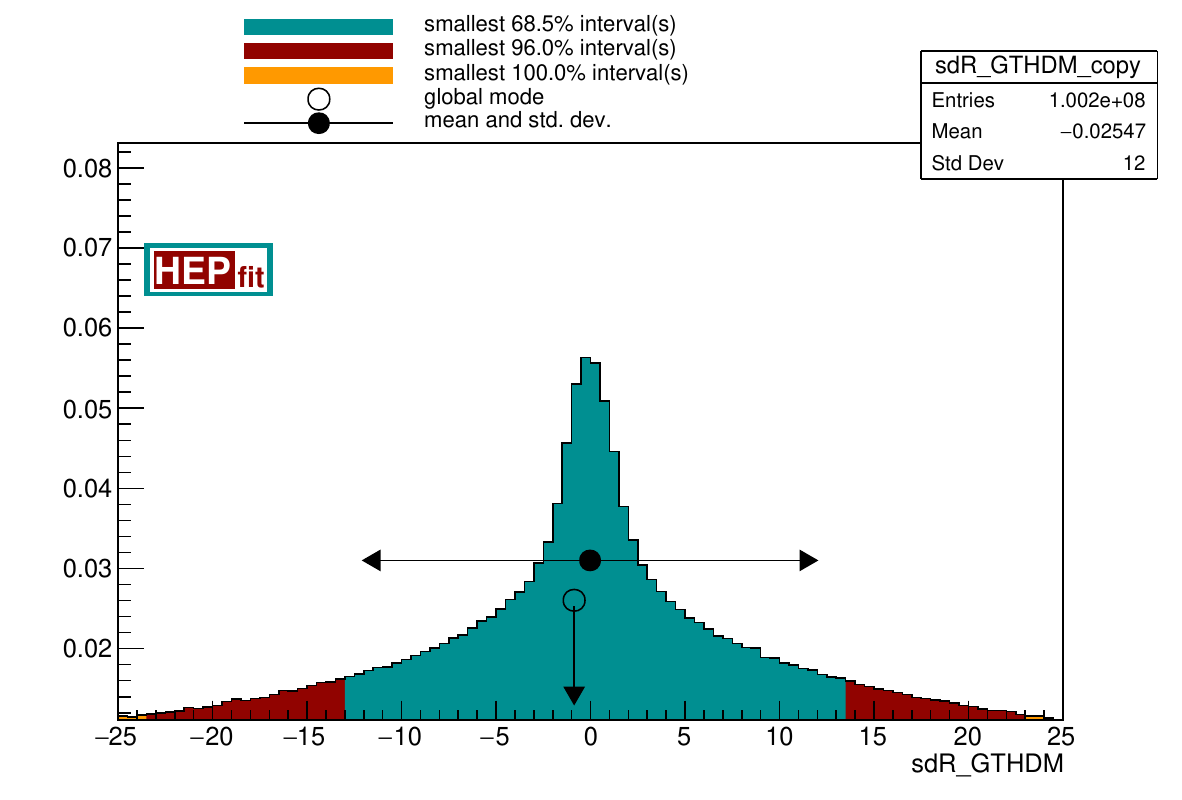}&
    \includegraphics[width=0.20\textwidth]{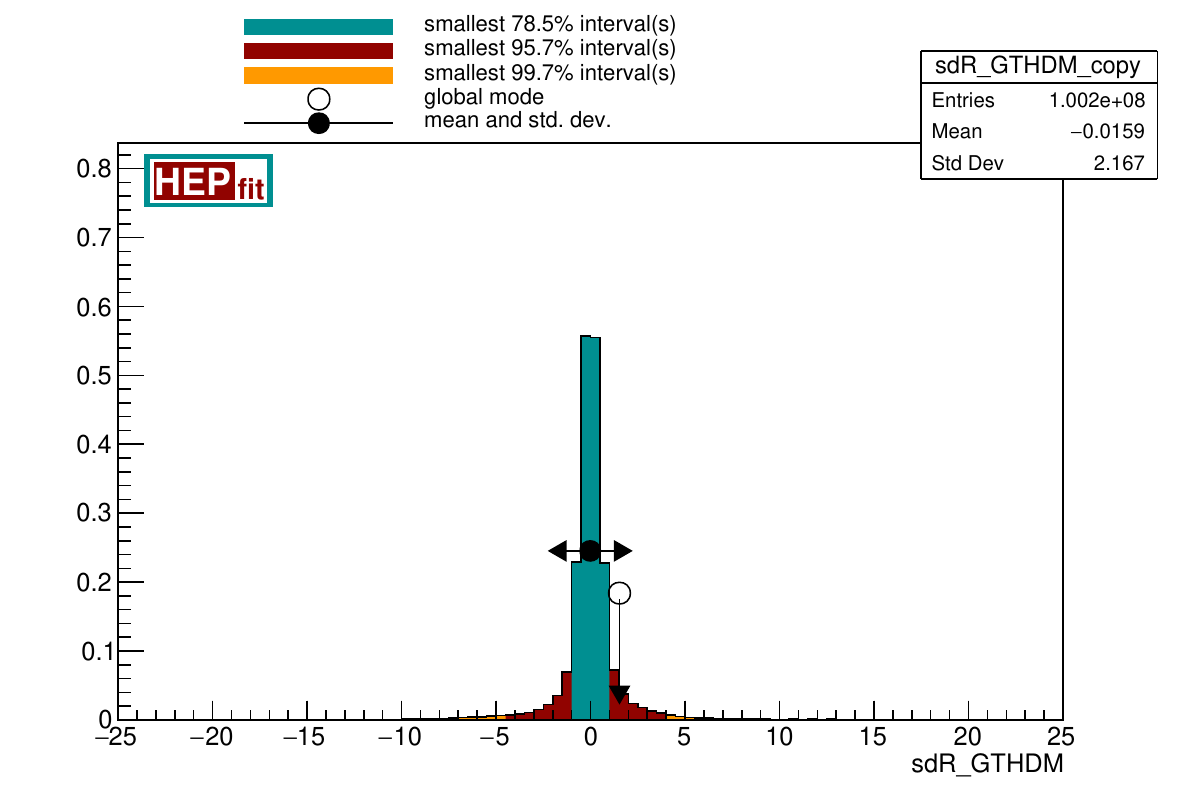}&
    \includegraphics[width=0.20\textwidth]{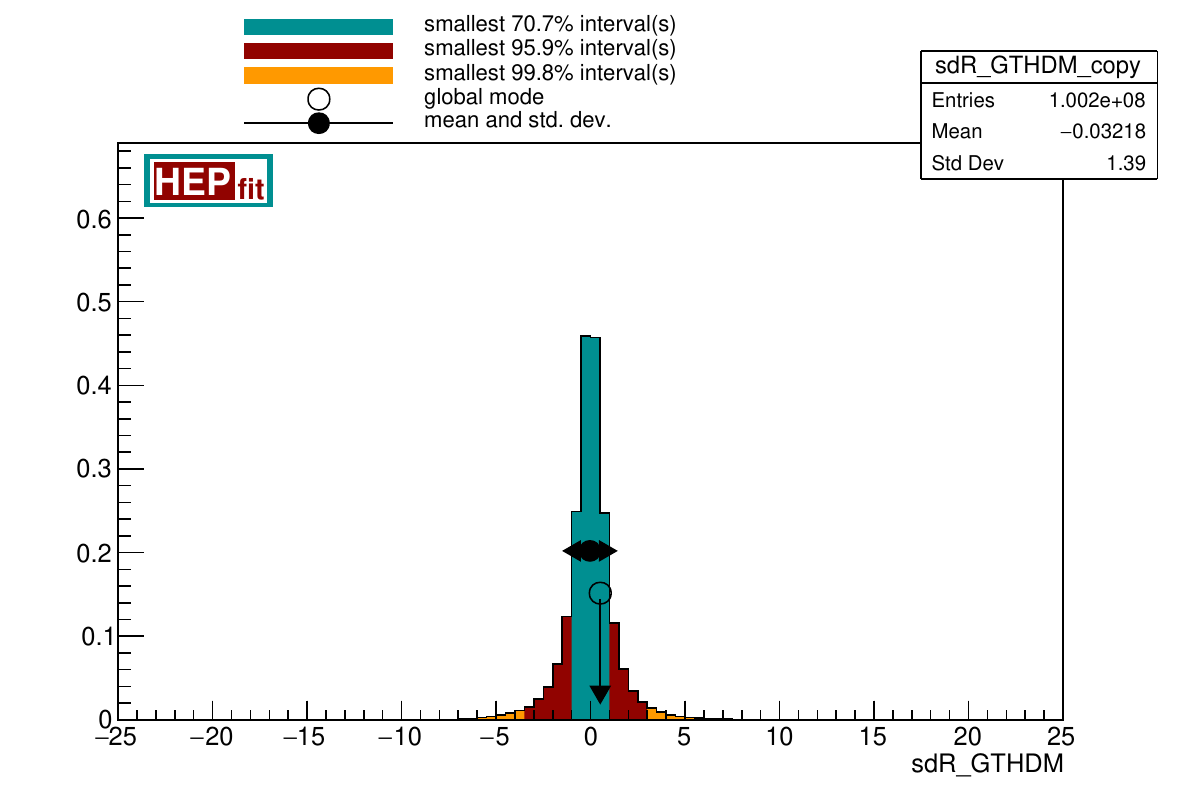}
    \\
    \includegraphics[width=0.20\textwidth]{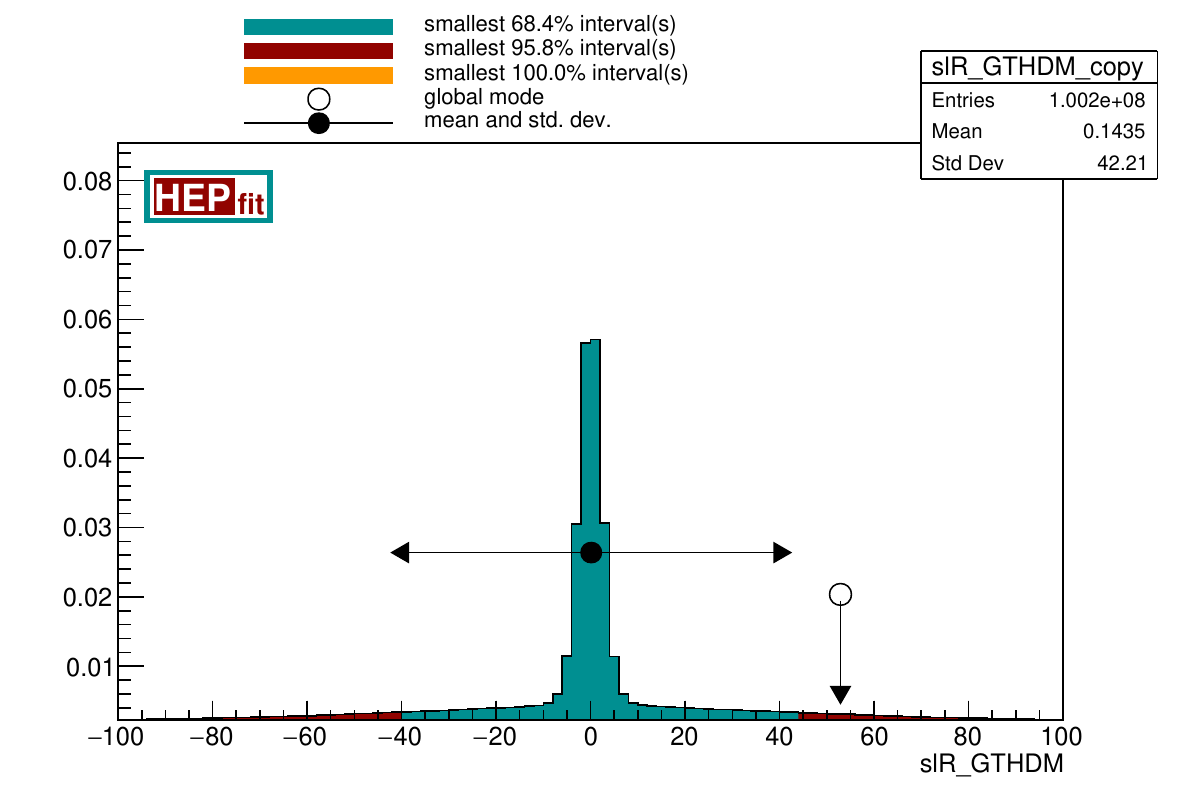}&
    \includegraphics[width=0.20\textwidth]{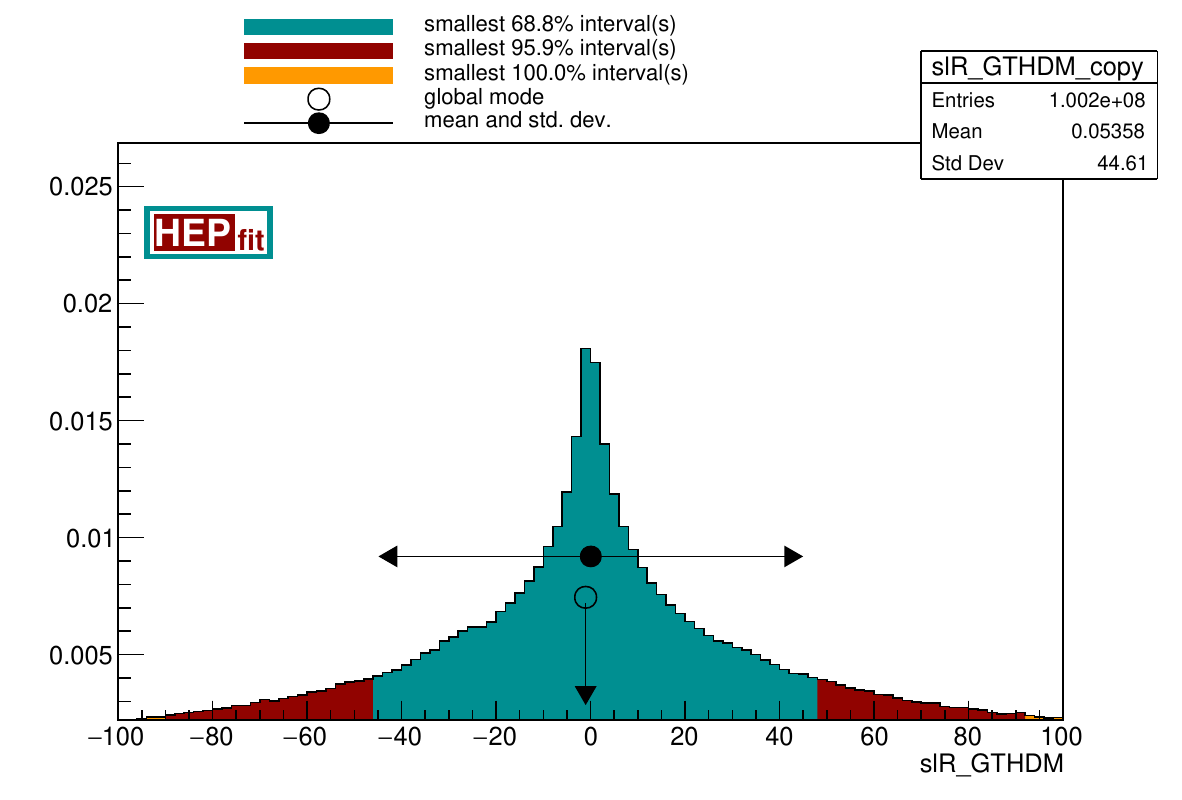}&
    \includegraphics[width=0.20\textwidth]{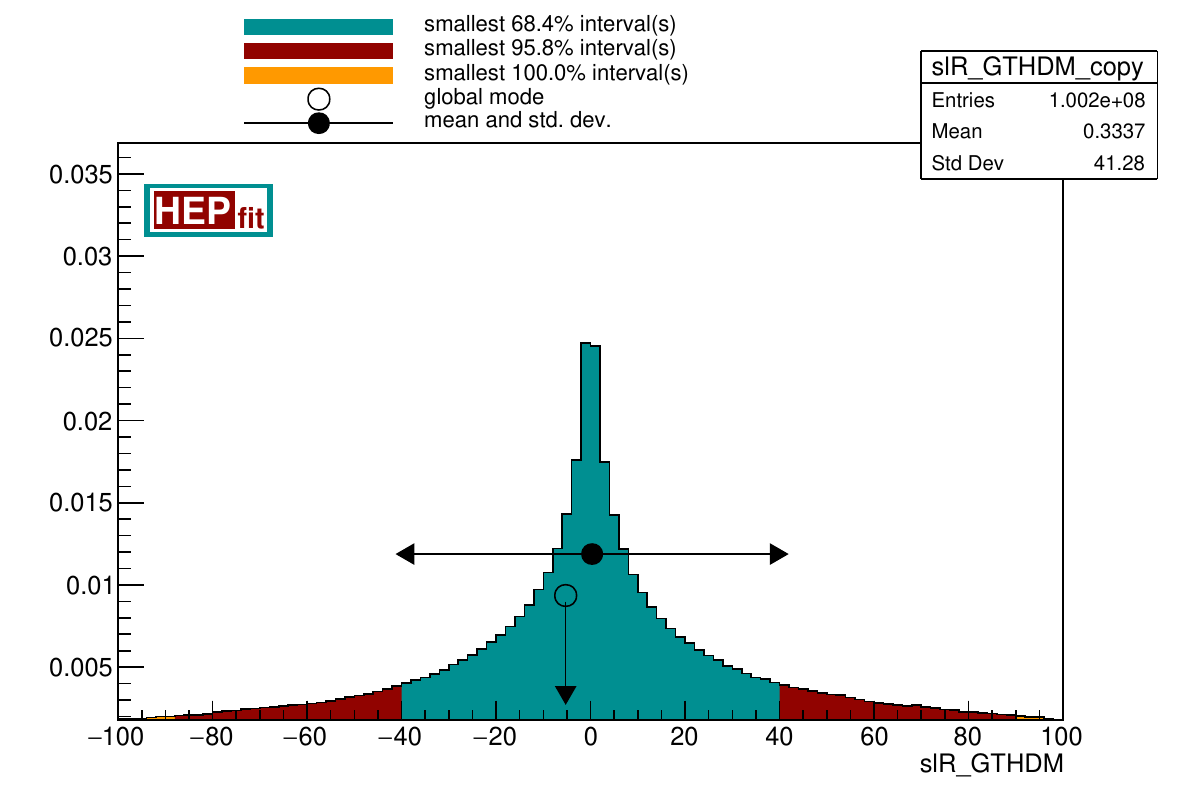}&
    \includegraphics[width=0.20\textwidth]{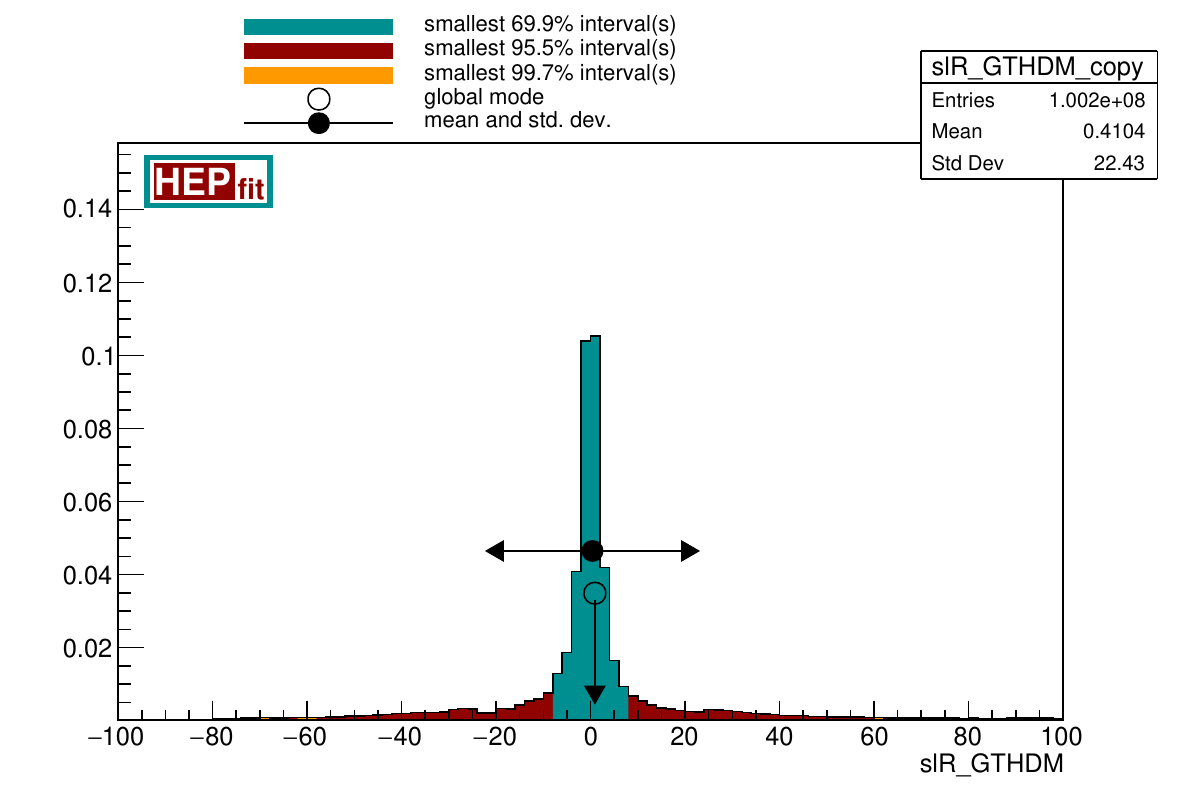}
    \\
    \hline\hline
  \end{tabular}
  }
  \caption{Posterior distributions of seven model parameters,
  $\lbrace M_A, M_H, M_{H^\pm}, \tilde\alpha, \varsigma_u, \varsigma_d, \varsigma_\ell \rbrace$, from top to bottom.
  Each column contains the results for each set of constraints analysed here (theory and EWPO are considered in every case): direct searches at LEP and LHC; Higgs signal strengths measured at LHC; flavour observables. The final column contains the global fit, where all data are simultaneously imposed.}
  \label{fig:fits}
\end{figure}

\clearpage

\section{Analyses of the Light-Pseudoscalar Scenario}
Identifying $h$ as the 125~GeV Higgs boson, the A2HDM can be parametrised by: three masses, $M_{A,H,H^{\pm}}$; a mixing angle, $\tilde\alpha$; three alignment coefficients, $\varsigma_{u,d,\ell}$; and three independent parameters from the potential, which can be chosen to be $\lbrace \lambda_2, \lambda_3, \lambda_7 \rbrace$. Efforts to study the parameters space of the A2HDM, how it gets affected by updates in experimental searches and which are the allowed regions that current and future colliders could inspect, have been published in the literature over the years, most recently in Ref.~\cite{Karan:2023kyj}. Both in the work done there and in the one we now present here, the analyses have been performed with the aid of \HEPfit{}~\cite{DeBlas:2019ehy}, a publicly available software which provides a Markov Chain Monte Carlo framework to determine the posterior distributions of model parameters when constrained by the likelihood distributions set by data, all within a Bayesian approach to statistical inference.

In the global analysis of Ref.~\cite{Karan:2023kyj} and in previous endeavours referenced therein, the extra scalars were always assumed to be heavier than the 125~GeV Higgs. For this new analysis, we set out to see whether the A2HDM is able to contain any scalar with a mass below that threshold. Focusing on the scenario where only the pseudoscalar is lighter than the Higgs boson, we have chosen the ranges presented in Tab.~\ref{tab:priors}, to be used as flat prior probability distributions. As per constraints, in the spirit of a global fit, we have implement in \HEPfit{} everything which might affect the A2HDM, namely for this analysis: purely theoretical considerations, such as the demands that the potential has a minimum, that this vacuum state is stable, that amplitudes computed in the model are unitary, and all perturbative expansions converge; electroweak precision observables (EWPO), including the so-called oblique parameters, and the ratio of decays of the $Z$ boson to bottom quarks and to hadrons; direct searches for neutral and charged scalars at LEP and LHC; the signal strengths of the Higgs boson measured at LHC; indirect searches such as the branching ratios of the decays $B\to X_s\gamma$, $B_s\to \mu\mu$, $B\to \tau \nu$, the mass difference of the $B_s$ system, flavour observables involving charmed mesons, $D_{(s)}\to \ell\nu$, and ratios of decays involving kaons and pions. For more details, including references to the data that is used, we refer to the dedicated section in Ref.~\cite{Karan:2024kgr}.

In order to investigate the effect different sets of constraints have on our model parameters, we perform four Bayesian fits: with theory constraints and EWPO always on, for a minimum of consistency of the sampled points, the first fit takes direct searches into account, the second looks at the signal strengths, in the third we turn to the flavour observables, and finally we put everything together for a global fit. The main results can be found in Fig.~\ref{fig:fits}. An immediate, and perhaps the most important, conclusion is that the global fit shows a viable region of parameter space, compatible with all the constraints we impose. Worth of remark is also the impact direct searches and signal strengths have on $M_A$ values below 62.5~GeV, where the channel of $h$ decaying into pairs of very light pseudoscalars is open, those limits basically ruling this out. While direct searches and flavour do not particularly constrain $\tilde\alpha$, the signal strengths seem force it to be almost compatible with zero, so as to render the 125~GeV Higgs very Standard-Model-like, as expected from what has been observed at the LHC so far. The alignment coefficients $\varsigma_f$ are also made to be quite small, as this helps branching ratios of the decays of scalar into pairs of fermions, present in much of direct searches and flavour observables, to stay below the current bounds. Concerning the two heavy scalars, every posterior indicates a preference for $M_H$ and $M_H{\pm}$ to always be, with the priors that were chosen for these four analyses, in a region within 200--400~GeV.

\section*{Acknowledgements}
This work has been supported by the Generalitat Valenciana (grant PROMETEO/2021/071),
by MCIN/AEI/10.13039/501100011033 (grant PID2020-114473GB-I00),
by the European Research Council under the European Union’s Horizon 2020 research and innovation programme (grant agreement 949451),
and by a Royal Society University Research Fellowship (grant URF/R1/201553).

\bibliographystyle{JHEP}
\bibliography{mybib}

\end{document}